\documentclass[aps,floatfix,showpacs,epsfig]{revtex4}
\usepackage{epsfig}
\usepackage{natbib}
\usepackage{graphicx}
\usepackage{amssymb}
\usepackage{amssymb,latexsym}
\usepackage[mathscr]{eucal}

\newcommand{\Lie}[0]{{\cal L}\, }
\newcommand{\grad}[0]{\nabla\!}
\newcommand{\R}{{\mathcal{R}}}
\def\th{{\widehat{\tau}}}
\def\rh{{\widehat{r}}}
\def\twoR{{\widetilde{\R}}}
\def\ls{{(\ell)}}
\def\ns{{(n)}}
\def\l{{\ell}}

\def\q{{\widetilde{q}}}
\def\K{{\widetilde{K}}}

\def\g{{\rm grav}}

\def\be{\begin{equation}}
\def\ee{\end{equation}}
\def\ba{\begin{eqnarray}}
\def\ea{\end{eqnarray}}
\begin{document}
\title{\large \bf A concrete anti-de Sitter black hole with dynamical horizon having toroidal cross-sections and its characteristics}

\author{Pouria Dadras}
\affiliation{Department of Physics, Sharif University of Technology,
Tehran, Iran}

\author{J. T. Firouzjaee}
\affiliation{School of Physics and School of Astronomy, Institute for Research in Fundamental Sciences (IPM), Tehran, Iran}
 \email{j.taghizadeh.f@ipm.ir}

\author{Reza Mansouri}
\affiliation{Department of Physics, Sharif University of Technology,
Tehran, Iran and \\
  School of Astronomy, Institute for Research in Fundamental Sciences (IPM), Tehran, Iran}
 \email{mansouri@ipm.ir}

\date{\today}

\begin{abstract}

  We propose a special solution of Einstein equations in the general Vaidya form representing a dynamical black hole having horizon cross sections
with toroidal topology. The concrete model enables us to study for the first time dynamical horizons with toroidal topology, its area law, and the question of
matter flux inside the horizon, without using a cut-and-paste technology to construct the solution.

\end{abstract}
\maketitle
\section{Introduction }

The topology of black hole horizons has been a matter of wide discussions in the past. Starting with the
Hawking's theorem, stating that each connected component of the event horizon of a
stationary black hole in four dimensional space time has the topology of a 2-sphere \cite{haw-book}, most
of the authors have been interested in non-dynamical asymptotically flat space times, excluding any kind of
cosmological black holes \cite{man}. Gannon was the first who opened the possibility of a torus topology for a black hole
horizon \cite{Gannon}, generalizing Hawking's theorem by replacing stationarity by some weaker assumptions.\\
\\
On the other hand, Chru´sciel and Wald \cite{wald} showed that each connected component of a cross-section
of the event horizon of a stationary asymptotically flat black hole must have spherical
topology. Jacobson and Venkataramani \cite{Jacobson} proved that, under certain conditions, the topology of the event
horizon of a four dimensional asymptotically flat black-hole space time must be 2-sphere. All theses studies
have been supported by the �topological censorship theorem� of Friedmann, Schleich and Witt, another statement
indicating the impossibility of non spherical horizons \cite{Friedman}. The theorem states that in a globally hyperbolic,
asymptotically flat spacetime, any two causal curves extending from past to future null infinity
are homotopic, meaning that a black hole with toroidal surface topology is a possible violation of topological
censorship theorem, as pointed out in \cite{Jacobson}. In fact, as was shown by Shapiro,
Teutolsky and Winicour \cite{Teutolsky}, a temporarily toroidal horizon can be formed in a gravitational collapse, in
a way consistent with the theorem.\\
Now, a concrete model of a black hole with toroidal horizon has been constructed by \cite{vanzo} for which the thermodynamics
and area law is also considered. Vanzo's model represents an isolated genus-one black hole in an asymptotically anti-de Sitter
space time; its extension to $d$ dimensions with a negative cosmological constant is given in \cite{Birmingham}.
Recently, toroidal and higher genus asymptotically AdS black holes have been put forward \cite{torus-collapse} through
gluing of some special metrics such as Lemaitre-Tolman-Bondi-like and McVittie solution to toroidal or higher genus asymptotically AdS black holes.
These pasted-manifolds, however, lack the dynamical features of the black hole we are interested in.\\

Our purpose is to construct a dynamical topological black hole in an asymptotically anti-de Sitter space time as a first step
towards a better understanding of cosmological black holes\cite{man} and their thermodynamics. For a more concrete definition of dynamical
horizons and their difference to the isolated ones we refer to \cite{ashtekar-03} and the references therein. A general formalism for the
area law in the case of dynamical black holes is also formulated there for the first time, including the possibility of
the toroidal cross section of the horizon and the corresponding black hole area law. Using a 3+1 decomposition and the Gauss-Bonnet
theorem, they also found a formalism which identifies the rate at which the radius of the cross-sections increases precisely according
to the matter flux and gravitational wave on the horizon with cross-sections having a spherical topology and strictly positive
cosmological constants. However, in the case of zero and negative cosmological constant, where the topology of horizon cross-sections
may be toroidal, their formalism is not conclusive.\\
In this paper we construct a non-stationary space-time which is asymptotically anti-de Sitter and represents a
dynamical horizon with toroidal cross-sections. An area law is then written for this dynamical black hole showing
an increase of the horizon in accordance with the second law of black hole thermodynamics. The matter flux, however, is
non-vanishing, although the total matter flux including the contribution from the cosmological constant vanishes. Our
dynamical horizon model with toroidal cross-sections reduce to that of Vanzo \cite{vanzo} with an isolated
horizon if the space time is stationary. \\

 We will follow in this paper definitions and notations used in \cite{ashtekar-03}. They introduced a
 local definition of horizon as a three dimensional manifold $H$ in space time which can be foliated by
 closed 2-dimensional surfaces $S$, assuming special characteristic of the expansion on each leaf.
 The space-time metric $g_{ab}$ has signature $(-,+,+,+)$ and
its covariant derivative operator will be denoted by $\nabla$. The Riemann
tensor is defined by $R_{abc}{}^d W_d := 2 \nabla_{[a}
\nabla_{b]}W_c$, the Ricci tensor by $R_{ab} := R_{acb}{}^c$, and the scalar curvature by $R := g^{ab} R_{ab}$. The
unit normal to $H$ will be denoted by $\th^a$; $g_{ab}\th^a\th^b
= -1$. The intrinsic metric and the extrinsic curvature of $H$ are
denoted by $q_{ab}:= g_{ab} + \th_a\th_b$ and
$K_{ab}:={q_a}^c{q_b}^d\grad_c\th_d$ respectively. $D$ is the covariant
derivative operator on $H$ compatible with $q_{ab}$, $\R_{ab}$ its
Ricci tensor and $\R$ its scalar curvature.  The unit space-like vector orthogonal to $S$ and
tangent to $H$ is denoted by $\rh^{\,a}$. Quantities intrinsic to $S$ will be generally written
with a tilde. Thus, the two-metric on $S$ is $\q_{ab}$ and the
extrinsic curvature of $S\subset H$ is
$\K_{ab}:=\widetilde{q}_a^{\,\,\,\,c}\widetilde{q}_b^{\,\,\,\,d}
D_c\rh_d$; the derivative operator on $(S, \q_{ab})$ is
$\widetilde{D}$ and its Ricci tensor is $\twoR_{ab}$. Finally, we can
fix the rescaling freedom in the choice of null normals
$\l^a:=\th^{\,a}+\rh^{\,a}$ and $n^a:=\th^{\,a}-\rh^{\,a}$ such that
$\ell^a n_a = -2$. The shear tensor $\sigma_{ab}$ of the null vector $l^a$ and its trace $\sigma$
are defined as usual.
A dynamical horizon is then defined as a three dimensional space-like sub-manifold
of space time such that it can be foliated with closed orientable two dimensional surfaces on which the
expansion $\Theta_{(\ell)}$ vanishes, and the expansion $\Theta_{(n)}$ of the other null normal is negative.
Section 2 is devoted to a short review of the area law formalism for dynamical horizons in different cases of
the cosmological constant. The area law for the dynamical horizon in a concrete metric with AdS  back ground
($\Lambda<0$) as an exact solution of the Einstein equations is discussed in section III. We then conclude in section IV.

\section{Area law for dynamical black holes }

Now, having the necessary definitions and notations, we may write the first consequence of the
above definition of a dynamical horizon, using the relations
$\Theta_\ls =0$ and $\Theta_\ns <0$, in the form

\be \label{area-in} \K = \tilde{q}^{ab} D_a \rh_b = \frac{1}{2}\,
\tilde{q}^{ab} \nabla_a (\ell_b -n_b) = - \frac{1}{2} \Theta_{(n)}
>0. \ee
Hence, the area $a_S$ of $S$ will increase monotonically along $\rh^{\,
a}$ by the change of the cross-sections, which is equivalent to the second law of black hole mechanics on $H$.
To obtain an explicit expression for the area change the authors in \cite{ashtekar-03} take
first two fixed cross-sections $S_1$ and $S_2$ of $H$, and then integrate the result on a
portion $\Delta H \subset H$ bounded by $S_1$ and $S_2$ with the corresponding
radii $R_1$ and $R_2$. Note that $R$ is the area radius of $S$ defined by $a_S = 4 \pi R^2$, $a_S$ being
the surface area of $S$ independent of its topology.
 Following result is then obtained using  Gauss-Bonnet theorem\cite{ashtekar-03}:

\be\label{ab1}  \mathcal{I}\, (R_2 - R_1) =  16\pi G \int_{\Delta H}
(T_{ab}- \frac{\Lambda}{8\pi G} g_{ab}) \th^{\,a}\xi_{(R)}^b\,d^3V +
\int_{\Delta H} N_R\left\{ |\sigma|^2 + 2|\zeta|^2\right\}\,d^3V, \ee
where $\mathcal{I}$ is the Euler characteristic of $S$; The scalar $|\zeta|$ is the length of the vector $\zeta^a=\tilde{q}^{ab}\hat{r}^c \nabla_b \ell_c$
and $\xi_{(R)}^a = N_R l^a$ where the lapse function is given by $N_R =|\partial R|$. The first term on the right hand side is usually called the matter flux, $ \mathcal{F}^{(R)}_{matter}$, and the second term is the gravitational wave
flux, $ \mathcal{F}^{(R)}_\g$. The discussion on the
topology of $S$ is now divided in three cases depending on the cosmological constant.

\begin{description}
\item[\emph{Case 1}:] $\Lambda > 0$. Since the stress energy
tensor $T_{ab}$ is assumed to satisfy the dominant energy condition,
the right hand side is manifestly positive definite. Due to the fact that the
area increases along $\rh^a$, we must have $R_2 - R_1
>0$. It then follows that $\mathcal{I}$ has to be positive. Therefore, the
closed, orientable 2-manifolds $S$ are necessarily
topologically 2-spheres and $\mathcal{I} = 8\pi$. Eq.(\ref{ab1}) now
gives a measure for the increase of the horizon cross-section area\cite{ashtekar-03}: %

\ba\label{ab2}  \frac{R_2-R_1}{2G} &=&  \int_{\Delta H} (T_{ab}-
\frac{\Lambda}{8\pi G}\, g_{ab}) \th^{\,a}\xi_{(R)}^b\,d^3V
\nonumber \\ &+& \frac{1}{16\pi G}\,\int_{\Delta H} N_R\left\{
|\sigma|^2 + 2|\zeta|^2\right\}\,d^3V\,. \ea
\item[\emph{Case 2}]: $\Lambda =0$. The right-hand side of eq.(\ref{ab1})
is necessarily non-negative. Hence, the topology of $S$ is either
that of a 2-sphere (if the right hand side is positive) or that of a
2-torus (if the right hand side vanishes).
The torus topology can occur if and only if $T_{ab}\ell^b$,
$\sigma_{ab}$ and $\zeta^a$ all vanish everywhere on $H$. Therefore, it may be
concluded that the scalar curvature $\twoR$ of $S$ must also vanish
on every cross-section. The 2-manifold $S$ then has to be a flat torus. Using the fact
that $H$ is space-like, we conclude that in this case ${\cal L}_n\, \Theta_{(\ell)} =0$ everywhere
on $H$. Thus, in this case the dynamical horizon cannot be a FOTH \cite{ashtekar-03}. Furthermore, since
 $\Theta_{(\ell)}, \sigma_{ab}$, and $R_{ab}\ell^b$ all vanish
on $H$, the Raychaudhuri equation now implies that ${\cal L}_\ell\,
\Theta_{(\ell)}$ also vanishes. \\
Note the following cases:

\begin{itemize}

\item In the case of torus topology the transition to the stationary case and
isolated horizon is not trivial\cite{ashtekar-03}. Given that in the stationary
case the topology can not be a torus \cite{haw-book}, a topology change is then
unavoidable \cite{horowitz-91}. The procedure used in \cite{ashtekar-03} to understand
the transition to the isolated horizon is based on the relation
 \be f\,\Lie_{n}\, \Theta_{(\ell)} =
\label{spacelike} -\sigma^2 - R_{ab}\ell^a\ell^b,
  \ee
where $f$ is the length of a space-like vector orthogonal to $S$ and tangent to $H$. Now, due to
the fact that in the case of torus topology we have $\Lie_{n}\, \Theta_{(\ell)} =0$, it can not be
concluded that $f=0$ which is a necessary condition for the isolated horizon with the sphere topology.\\

\item The familiar matter flux as defined in (\ref{ab1}) does vanish in the case of the torus topology. However, by
changing the definition of $\xi_{(R)}$ in (\ref{ab1}) to $\xi_{(R)}^b = c n^b$, where $c$ is
an appropriate coefficient related to $f$, we may arrive at a non-vanishing matter flux. That this
is not always the case, we will see on hand a concrete example in the following section. We may also note
that the positivity of the extrinsic curvature of $S$ along $r^a$ in the case of torus topology does not
necessarily means an area increase. This is due to the vanishing of the left hand side of (\ref{ab1}).
\end{itemize}
\item[\emph{Case 3}:] $\Lambda <0$. In this case there is no control on
the sign of the right-hand side of eq.(\ref{ab1}). Hence, any topology is permissible. Stationary
solutions with quite general topologies are known for black holes which are locally
asymptotically anti-de Sitter\cite{vanzo}. Event horizons of these solutions are
potential asymptotic states of these dynamical horizons in the
distant future.
\end{description}

\section{Area law for a dynamical black hole in AdS back ground }

We are interested in a solution of Einstein equations with negative cosmological constant representing a black hole
with a dynamical horizon having torodial cross-sections. So far we have not found any exact solution of Einstein equations
having these features and not constructed through a cut and paste technology. Solutions produced by cut and paste technology
do not represent a genuine dynamical black hole due to the build-in freezing of the matching hypersurface. We propose
then a solution in the general Vaidya form:

\be ds^{2} = -f(v,r) dv^2 + 2dvdr +
r^2 (d\theta^2 + d\phi^2),  \ee
with the arbitrary function $f$ of coordinates $v$ and $r$, where $v$ is the advanced
time coordinate with $-\infty<v<\infty$, $r$ is the radial coordinate
with $0<r<\infty$, and $\theta,\phi$ are coordinates describing the two-dimensional
zero-curvature space generated by the
two-dimensional commutative Lie group $G_2$ of isometries \cite{vanzo}.\\
The black-hole apparent horizon is space-like and located at
\be \label{eq:cubic} f(v,r) := 0 \, . \ee
The expansions of the corresponding null normals are
\be \Theta_{\ell} = \frac{f}{r}, \qquad \textrm{and} \qquad
\Theta_{n} = -\frac{4}{r} \, .\ee
Note that $\Theta_{n}$ is always negative and
$\Theta_{\ell}$ vanishes precisely at the horizon, as required by a dynamical horizon.
The unit normal to the horizon is given by
\be \hat{\tau}_a =
\frac{1}{\sqrt{|2\dot{f}f^\prime|}}[\dot{f}, f^\prime] \qquad \textrm{and} \qquad \hat{\tau}^a =
\frac{1}{\sqrt{|2\dot{f}f^\prime|}}[f^\prime, \dot{f}] \, . \ee
The constant $r$ surfaces are the preferred cross-sections of the
horizon and the unit space-like normal $\hat{r}^a$ to these cross
sections is given by
\be \hat{r}_a =
\frac{1}{\sqrt{|2\dot{f}f^\prime|}}[-\dot{f},
f^\prime] \qquad \textrm{and} \qquad \hat{r}^a =
\frac{1}{\sqrt{|2\dot{f}f^\prime|}} [f^\prime,- \dot{f}].
\ee
The properly rescaled null normals are then given by
\be \ell^a = \frac{2|f^\prime|}{\sqrt{|2\dot{f}f^\prime|}}(1,0,0,0)  \qquad \textrm{and} \qquad
n^a = \frac{2\dot{f}}{\sqrt{|2\dot{f}f^\prime|}}(0,1,0,0)  \, . \ee
The lapse function corresponding to the radial coordinate $r$, which
in this case is identical to the area radius, is given by
\be N_r = \left|\frac{\dot{f}}{2f^\prime}\right|^{1/2}. \ee
Thus the properly rescaled vector field corresponding to the
radial coordinate $r$ is $\xi_{(R=r)}^a = N_r \ell^a = (\partial /\partial v)^a$.
\\
Now, take the following special solution with toroidal horizon configurations suggested in \cite{lemos97}
\begin{equation}
ds^2 = - \left(\alpha^2r^2-\frac{\beta m(v)}{r}\right) dv^2 + 2 dv dr +
r^2 (d\theta^2 + d\phi^2),
                         \label{eq:2.2}
\end{equation}
where $0\leq\theta<2\pi$, $0\leq\phi<2\pi$. The corresponding energy-momentum tensor is then given by
\begin{eqnarray}
&T_{ab} = \frac{\beta}{8\pi r^2} \frac{dm(v)}{dv} k_ak_b, &\nonumber \\
&  \quad k_a = - \delta^v_a, \quad k_ak^a = 0, &
                        \label{eq:2.3}
\end{eqnarray}
where $\alpha \equiv \sqrt{\frac{-\Lambda}{3}}$, $\beta=q/\alpha$, and $m(v)$ is the Misner-Sharp mass.
In these coordinates, lines with $v = $constant represent incoming radial
null vectors having tangent vectors in the form $k^a=(0,-1,0,0)$,
or $k_a = (-1,0,0,0)$. The energy momentum tensor depends, in general, on $\Lambda$ and
diverges as $\Lambda \rightarrow 0$. To assume a toroidal cross-section and apply the eq.(\ref{ab1}), we
therefore ask for $\beta$ to be independent of $\Lambda$, leading to $q=\frac{2\alpha}{\pi}$.

Now, noting that the energy-density of the radiation is
$\epsilon = \frac{1}{4\pi^2 r^2}\frac{dm}{dv}$, one sees that
the weak energy condition for the radiation is satisfied whenever
$\frac{dm}{dv}\geq0$, i.e., the radiation is imploding.\\
The apparent horizon surface is now defined by
\begin{equation}
qm|_{AH}=\alpha^3 r|_{AH}^3.
\end{equation}

Having specified $f$, and noting that $ \dot{f}=-\frac{q \dot{m}}{\alpha r}$ and $ f'= \frac{2 \alpha^3 r^3+qm}{\alpha r^2}$, we are
able to calculate the gravitational flux leading to

\begin{equation}
\mathcal{F}^{(R)}_\g = \frac{1}{16\pi G}\,\int_{\Delta H} N_R\left\{ |\sigma|^2 + 2|\zeta|^2\right\}\,d^3V=0.
\end{equation}
Therefore, from the equation (\ref{ab1}) we conclude that in our model of dynamical black hole with the toroidal topology the total matter flux term
including the $\Lambda$-fluid across the horizon has to be zero, \textit{i.e}. $ \mathcal{F}^{(R)}_{total} =0 $. This means that the ordinary matter flux
$ \mathcal{F}^{(R)}_{ordi}= \int_{\Delta H}
T_{ab}\th^{\,a}\xi_{(R)}^b\,d^3V$, equals to the corresponding \textit{$\Lambda $-fluid} term, i.e. $\mathcal{F}^{(R)}_{\Lambda}=\int_{\Delta H} \frac{\Lambda}{8\pi G} g_{ab}) \th^{\,a}\xi_{(R)}^b\,d^3V$.\\
Therefore, in the general case of non-stationary metric where $m(v)$ is everywhere time dependent, the horizon is dynamical, the total matter flux
is vanishing while the ordinary matter flux is non-zero, and the horizon is space-like. Now, let us differentiate two special cases:

\begin{itemize}
\item \textit{$m$ is constant everywhere} : In this case, the horizon is isolated.  The following coordinate transformations
will then give us the stationary metric suggested by Vanzo \cite{vanzo}:
\begin{equation}
t=v-\int \frac{dr}{\left(\alpha^2r^2-\frac{q m}{\alpha r}\right)}.
\end{equation}
Therefore, all results stated in \cite{vanzo} for this metric including the formula for the black hole area law are applicable here.

\item \textit{$m$ is constant on the horizon ($m|_{AH}=constant$)} : \\
In this limiting case we have $\dot{m}|_{AH}=0$, i.e. the matter flux in addition to the total flux is zero, and $\th^{\,a}$ will be
null. Thus, the horizon, being null now, has a constant surface. It is therefore an isolated horizon although the metric is non-stationary.

\end{itemize}

\section{Conclusion}

We have constructed a dynamical black hole having toroidal topology in its cross-sections within an asymptotically anti-de Sitter
space time as an exact solution of Einstein equations not produced by any cut-and-paste technology. The area law is written out and
it has been shown that the total matter flux, including the $\Lambda$-matter, is zero while the ordinary matter flux is non-vanishing.
This model for the first time, exemplifies the existence of dynamical horizons with toroidal topology. The vanishing of the
total matter flux may be just a feature of the concrete model we have proposed. The model leads in a limiting case to an
isolated horizon. Assuming our general metric to be stationary it reduces to the Vanzo metric \cite{vanzo}.

\section{ACKNOWLEDGMENT}
We would like to thank Abhay Ashtekar for suggesting the problem.

\end{document}